\RequirePackage{ifpdf}
\ifpdf 
\documentclass[pdftex]{sigma}
\else
\documentclass{sigma}
\fi

\renewcommand{\vec}{\boldsymbol}
\newcommand{\rmd}{\mathrm{d}}

\newcommand{\Eset}{\mathbb{E}}

\newcommand{\cK}{\mathcal{K}}
\newcommand{\cM}{\mathcal{M}}
\usepackage{yfonts}
\newfont{\gothic}{eufm10 at 12pt}

\numberwithin{proposition}{section}
\numberwithin{remark}{section}

\begin{document}

\renewcommand{\PaperNumber}{***}

\FirstPageHeading

\ShortArticleName{The Smorodinsky-Winternitz potential}

\ArticleName{The Smorodinsky-Winternitz potential revisited}

\Author{Roman G. Smirnov~$^\dag$ and Amelia Yzaguirre~$^\ddag$}

\AuthorNameForHeading{R.G. Smirnov and A. Yzaguirre}

\Address{$^\dag$~Department of Mathematics and Statistics, Dalhousie University, Halifax, NS, CANADA B3H 3J5} 
\EmailD{\href{mailto:smirnov@mathstat.dal.ca}{smirnov@mathstat.dal.ca}} 
\URLaddressD{\url{http://www.mathstat.dal.ca/~smirnov/}} 

\Address{$^\ddag$~Department of Mathematics and Statistics, Dalhousie University, Halifax, NS, CANADA B3H 3J5}
\EmailD{\href{mailto:ameliay@mathstat.dal.ca}{ameliay@mathstat.dal.ca}} 


\ArticleDates{Received ???, in final form ????; Published online ????}

\Abstract{We employ joint invariants of Killing two-tensors defined in the Euclidean plane to characterize the Smorodinsky-Winternitz potential and explain the geometric meaning of its arbitrary parameters. In addition, we verify for which values of the arbitrary parameter $k$ the Tremblay-Turbiner-Winternitz potential is  multi-separable. }


\Classification{34J14; 37J15; 37J30; 51P05; 70S10; 70H20} 

\section{Introduction}

Solving both classical and quantum Hamiltonian systems follows the pattern: the more ``structure" (i.e., symmetries, first integrals, etc.) a Hamiltonian system has the more likely it can be solved explicitly. Superintegrable systems are very rich in structure, which makes it possible to solve them explicitly both analytically and algebraically. In this paper we investigate maximally-superintegrable Hamiltonian systems of classical mechanics defined by natural Hamiltonians which admit quadratic in the momenta first integrals  of motion. Recall that a Hamiltonian system with $n$ degrees of freedom is said to be maximally-superintegrable if it admits $2n-1$ global,  functionally independent, single-valued integrals of motion that do not depend explicitly on time. In what follows we also assume that such systems admit quadratic in the momenta first integrals of motion, which under some additional conditions afford orthogonal separation of variables in the associated Hamilton-Jacobi equation. In fact, such superintegrable systems can be solved via orthogonal separation of variables in more than one way, which is the reason they are called multi-separable. This particular  class of  superintegrable systems includes such important examples as  the Kepler potential, Calogero-Moser model, harmonic oscillator and others. 

Many efforts have been made over the years to classify superintegrable Hamiltonian systems defined in spaces of constant curvature of the type described  above. Formally, it was initiated  in 1965 by  Fri\v{s} {\em et al} who  produced \cite{FMSUW65} a list of superintegrable potentials of this type defined in Euclidean plane. In 1990 Evans \cite{evans90} extended this classification  to  superintegrable potentials defined in Euclidean 3D-space. More recently, several authors have launched the project of classification of superintegrable systems, based on the algebraic properties of the   symmetry algebra associated with a given superintegrable system. This idea has proven to be quite fruitful since the symmetry algebra of a superintegrable Hamiltonian system has a rich  structure, unlike the symmetry  algebra of a  completely integrable Hamiltonian system admitting only $n$ first integrals in involution,  whose symmetry algebra is   merely abelian (for more details refer to Kalnins {\em et al} \cite{KKM12} and the relevant references therein). 

Another approach to the problem of classification of superintegrable systems of the type described above defined in the Euclidean plane  has been realized by Adlam {\em et al} \cite{AMS08}    within the framework of the invariant theory of Killing tensors (ITKT) (see \cite{adlam05, AMS08, cdm06,  ccm09, cochran11, CMS11, horwood08, horwood07, HMS05, HMS09, MST02, smirnov08, smirnov11, SY04, WF65,  SY04, yue05} and the relevant references therein for more details). More specifically, the authors derived an invariant characterization of  the superintegrable potential of the Kepler problem, using  invariants and joint-invariants of Killing two-tensors that determined first integrals admitted by the system in question. 

In this work we extend some of the ideas employed in \cite{AMS08} to characterize, using  joint invariants  of Killing two-tensors \cite{SY04} defined in the Euclidean plane, the well-known Smorodinsky-Winternitz (SW) potential and explain the geometric meaning of its arbitary parameters. We also verify that  the important integrable perturbation of the Smorodinsky-Winternitz potential known as Tremblay-Turbiner-Winternitz (TTW) potential loses the property of multi-separability for all but one value of the arbitrary parameter $k$ (refer to the formula (\ref{sw})).

\section{Superintegrable, multi-separable systems in Euclidean plane} 
\label{super}

Let $(\cM, g)$ be an $n$-dimensional  pseudo-Riemannian space of constant curvature. Consider a Hamiltonian system defined in the cotangent bundle $T^*\cM$ by the following natural Hamiltonian function

\begin{equation}
H(\vec{q},\vec{p}) = \frac{1}{2}g^{ij}(\vec{q})p_ip_j + V(\vec{q}), \quad i,j = 1,\ldots ,  n, \label{h1}
\end{equation}
where $\vec{q} = (q^1,\ldots, q^n)$, $\vec{p} = (p_1, \ldots, p_n)$ are the corresponding canonical  coordinates, $g^{ij}$ denotes the components of the metric tensor, and $V(\vec{q})$ is the potential function. In what follows we will study the Hamiltonian systems defined by (\ref{h1}) that admit $2n-1$ first integrals of motion, given by 
 single-valued, functionally independent, globally defined  functions
\begin{equation}
F_1= H,  F_2, \ldots , F_{n}, F_{n+1}, \ldots, F_{2n-1} \in {\cal F}(T^*\cM). \label{fi}
\end{equation}
that  enjoy the property 
\begin{equation}
\{F_i, F_j\} = \{F_k,F_{\ell}\} = 0, \quad i,j = 1,\ldots, n;\, k,\ell = 1, n+1, \ldots, 2n-1, \label{inv}
\end{equation} 
where $\{ \cdot, \cdot\}$ denotes the corresponding (canonical) Poisson bracket defined in ${\cal F}(T^*\cM)$. A Hamiltonian system described above is said to be a {\em maximally-superintegrable} Hamiltonian system defined by a natural Hamiltonian (\ref{h1}). Furthermore, let the first integrals (\ref{fi}) be quadratic in the momenta $\vec{p}$ that is of the form
\begin{equation}
F_{\ell}(\vec{q},\vec{p}) = K_{\ell}^{ij}(\vec{q})p_ip_j + U_{\ell}(\vec{q}),   \label{fiq}
\end{equation}
where $\ell= 1,\ldots, 2n-1$, $i,j = 1,\ldots, n$. As is well known, the condition $\{H, F_{\ell}\}= 0$ implies for each $\ell = 1,\ldots, 2n-1$ that
\begin{equation}
[K_{\ell}, g] = 0, \label{kte}
\end{equation}
where $[\cdot, \cdot]$ denotes the Schouten bracket \cite{Sch} and
\begin{equation}
\rmd(\hat{K}_{\ell}\rmd V) = 0, \label{cc}
\end{equation}
where  $\hat{K}_{\ell} = K_{\ell}g^{-1}$. Equation (\ref{kte}) is called the {\em Killing tensor equation}, implying that $K^{ij}_{\ell}$ are the components of a Killing two-tensor $K_{\ell}$, while Equation (\ref{cc}) is said to be the correspodning {\em compatibility condition} (known also as the Bertrand-Darboux equation(s)). 

Naturally, one of the most important problems that concerns this type of superintegrable Hamiltonian systems is the problem of classification. Thus far it has been shown that it can be approached from different directions by taking into  account the algebraic, analytic and geometric properties of  superintegrable potentials in question and their corresponding   first integrals of motion. 

First of all, different superintegrable potentials can be derived once the Killing tensors defining the quadratic part of first integrals of motion (\ref{fiq}) have been classified (see Section \ref{itkt}) by solving the compatibility condition (\ref{cc}) for various sets of Killing two-tensors that determine quadratic in the momenta first integrals of motions (\ref{fiq}). This approach has been successfully employed, for example,  to derive lists of superintegrable potentials defined in the Euclidean plane and 3D-space (see \cite{FMSUW65} and \cite{evans90} respectively)

Next, as it has been shown by various authors in recent years (see \cite{KKM12} and the relevant references therein)  one can use in this context the structrue algebra of first integrals associated with a given superintegrable Hamiltonian system in view of the fact that it is not abelian (\ref{inv}). More specifically, for a given superintegrable Hamiltonian potential (\ref{h1}) that admits $2n-1$ first integrals of motion (\ref{fi}) one can determine the order at which the algebra of first integrals closes with respect to the Poisson bracket, which is an algebraic characteristic of the superintegrable potential in question. 

Alternatively, one can make use of the invariant properties of the Killing two-tensors that appear in (\ref{fiq}) by computing their joint invariants \cite{SY04} and characterizing the associated superintegrable potential accordingly. This approach has been employed by Adlam {\em et al} \cite{AMS08} to derive an invariant characterization  of the Kepler potential defined in the Euclidean plane $\Eset^2$. 

In this work our goal is twofold. First, we wish to characterize the {\em Smorodinsky-Winternitz (SW) superintegrable system} defined in $\Eset^2$  in terms of Cartesian coordintes $(q^1, q^2) = (x,y)$ by the following natural Hamiltonian function:

\begin{equation}
H = \frac{1}{2}(p_x^2 + p_y^2) - \omega (x^2 + y^2) + \frac{\alpha}{x^2} + \frac{\beta}{y^2}. \label{sw}
\end{equation}

This system is an integrable perturbation of the harmonic oscillator; its potential was derived in \cite{FMSUW65} in the framework of the problem of classification of superintegrable potentials defined in $\Eset^2$ that admitted two quadratic in the momenta first integrals of motions (\ref{fiq}), using the first approach reviewed above. In what follows we target to derive an invariant characterization of the potential of (\ref{sw}) in terms of the joint invariants of the associated Killing two tensors and explain the meaning of the arbitrary parameters $\omega$, $\alpha$ and $\beta$ from this viewpoint. 

Our second goal is to investigate a generalization of the SW potential given by (\ref{sw}) and known as the {\em Tremblay-Turbiner-Winternitz (TTW) potential}. The TTW  potential was obtained by Trembley {\em et al} \cite{TTW09, TTW10} by rewriting the quantum analogue of the natural Hamiltonian (\ref{sw}) in  polar coordinates $(r,\theta)$ and then replacing $\theta $ with $k\theta $, thus introducing a new parameter $k$. The resulting natural Hamiltonian 

\begin{equation}
H = p_r^2 + \frac{1}{r} + \frac{1}{r^2}p_{\theta}^2 -\omega r^2 + \frac{\alpha}{r^2 \cos^2 k\theta} + \frac{\beta}{r^2 \sin^2 k\theta} \label{ttw} 
\end{equation} 
is characterized by a potential that preserves some of the properties of the SW potential determined by (\ref{sw}). In particular, it is obviously still  separable in polar coordinates. The big question is whether or not the system is still superintegrable. The authors conjectured that the system defined by the potential of  (\ref{ttw}) was classically and quantum  superintegrable for all rational values of $k$. This conjecture has had a major impact on the development of the theory of superintegrable systems, resulting in a host of researchers  trying to prove it (for more details, see \cite{KKM12} and the relevant references therein). In this work we will show that the Hamiltonian system defined by (\ref{ttw}) is superintegrable in the sense of admitting quadratic first integrals of motion iff $k = \pm 1$, that is when it is of the form defined by (\ref{sw}).

\section{Invariant theory of Killing tensors in Euclidean plane}
\label{itkt}

Recall that the invariant theory of Killing tensors (ITKT) has been developed to study intrinsic properties of Killing tensors defined in pseudo-Riemannian spaces of constant curvature modulo the action of the corresponding isometry groups. This arrangement  makes it an analog of the classical theory of invariants of homogenous polynomials which are studied within the framework of the latter theory under the action of the general linear group or its subgroups (see Olver \cite{olver99} and the relevant references therein for more details). 

The origins of the theory can be traced back to the 1965 paper by Winternitz and Fri\v{s}  \cite{WF65} in which the authors studied second order symmetries of the Laplace operator defined in $\Eset^2$  modulo the action of the special Euclidean group $SE(2)$. This problem is the quantum analog of the corresponding problem in classical mechanics involving the Killing two-tensors defined in $\Eset^2$ under the action of $SE(2)$, which was independently solved in 2002 by McLenaghan {\em et al} \cite{MST02}. In the latter case the authors not only computed the invariants of the group action $SE(2) \circlearrowright \cK^2(\Eset^2)$, where $\cK^2 (\Eset^2)$ denotes the vector space of Killing two-tensors in $\Eset^2$, and used them to discriminate between the different orbits of the group action, but also computed in each case the corresponding moving frames map \cite{FO98, FO99, olver99}, as well as applied the resulting   algorithm to solve Hamiltonian systems defined on $\Eset^2$ that arised in physics via separation of variables. These results laid the groundwork that allowed a number of authors to successfully employ this approach in a more general setting to study Killing tensors defined in spaces of constant curvature (for more details, see \cite{adlam05, AMS08, cdm06,  ccm09, cochran11, CMS11, horwood08, horwood07, HMS05, HMS09, MST02, smirnov08, smirnov11, SY04, WF65,  SY04, yue05} and the relevant references therein). 

In what follows our notations are compatible with those adopted in \cite{AMS08}. In this section we adapt the ITKT to derive in Section \ref{icsw} an invariant characterization of the Smorodinsky-Winternitz potential given by (\ref{sw}). Since our goal is to characterize a superintegrable system defined in $\Eset^2$ that admit two distinct quadratic first integrals (other than the Hamiltonian), we first discuss how the theory works when there are two Killing two-tensors that define the first integrals of the form (\ref{fiq}).    Consider the product space $P$ defined by
\begin{equation}
P:= \cK^2(\Eset^2)\times \cK^2(\Eset^2), \label{ps}
\end{equation}
in which each of the two factors is given by the following formula:
\begin{equation}
\begin{array}{rcl}
K & = & \displaystyle (\beta_1 + 2\beta_4y + \beta_6y^2)\partial_1
\odot  \partial_1 \\ 
& & + \displaystyle (\beta_3 - \beta_4x -
 \beta_5y - \beta_6xy)\partial_1\odot \partial_2 \\ 
& & + \displaystyle (\beta_2 + 2\beta_5x+\beta_6x^2) \partial_2\odot
\partial_2,
\end{array}
\label{gKt}
\end{equation}
representing the solution space of the Killing tensor equation (\ref{kte}) for the case $\cM = \Eset^2$, where the arbitrary six parameters $\beta_i$, $i=1,\ldots, 6$ are the corresponding constants of integration and $\partial_1 = \frac{\partial}{\partial x}$, $\partial_2 = \frac{\partial}{\partial y}$. Note $\dim \cK^2(\Eset^2) = 6$. 

 Recall that the action $SE(2) \circlearrowright \Eset^2$ given by  
\begin{equation}
\begin{array}{l}
\tilde{x} = x\cos p_3 - y\sin p_3 + p_1, \\ 
\tilde{y} = x\sin p_3 + y\cos p_3 + p_2 
\end{array}
\label{A1}
\end{equation}
induces the corresponding action $SE(2)\circlearrowright P$: 
 \begin{equation}
\label{JA}
\begin{array}{rcl}
\tilde{\alpha_1} &=& \alpha_1\cos^2p_3 - 2\alpha_3\cos p_3\sin p_3 + \alpha_2\sin^2p_3 - 2p_2\alpha_4\cos p_3 
\\
                  & & - 2p_2\alpha_5\sin p_3 + \alpha_6p_2^2, \\ 
\tilde{\alpha_2} & =& \alpha_1\sin^2p_3 - 2\alpha_3\cos p_3\sin p_3 +
\alpha_2\cos^2p_3 - 2p_1\alpha_5\cos p_3  \\ 
 & &+ 2p_1\alpha_4\sin p_3  + \alpha_6p_1^2,\\
\tilde{\alpha_3} & = & (\alpha_1-\alpha_2)\sin p_3\cos p_3 + \alpha_3(\cos^2p_3 - \sin^2p_3) 
\\ 
 & &  + (p_1\alpha_4 + p_2\alpha_5)\cos p_3 + (p_1\alpha_5 - p_2\alpha_4)\sin p_3 - \alpha_6p_1p_2, \\
 \tilde{\alpha_4} & = & \alpha_4\cos p_3 + \alpha_5\sin p_3 - \alpha_6p_2, \\
 \tilde{\alpha_5} & = & \alpha_5\cos p_3 - \alpha_4\sin p_3 - \alpha_6p_1,\\
 \tilde{\alpha_6} &=& \alpha_6,\\
\tilde{\beta_1} &=& \beta_1\cos^2p_3 - 2\beta_3\cos p_3\sin p_3 + \beta_2\sin^2p_3 - 2p_2\beta_4\cos p_3 
\\
                  & & - 2p_2\beta_5\sin p_3 + \beta_6p_2^2, \\ 
\tilde{\beta_2} & =& \beta_1\sin^2p_3 - 2\beta_3\cos p_3\sin p_3 +
\beta_2\cos^2p_3 - 2p_1\beta_5\cos p_3  \\ 
 & &+ 2p_1\beta_4\sin p_3  + \beta_6p_1^2,\\
\tilde{\beta_3} & = & (\beta_1-\beta_2)\sin p_3\cos p_3 + \beta_3(\cos^2p_3 - \sin^2p_3) 
\\ 
 & &  + (p_1\beta_4 + p_2\beta_5)\cos p_3 + (p_1\beta_5 - p_2\beta_4)\sin p_3 - \beta_6p_1p_2, \\
 \tilde{\beta_4} & = & \beta_4\cos p_3 + \beta_5\sin p_3 - \beta_6p_2, \\
 \tilde{\beta_5} & = & \beta_5\cos p_3 - \beta_4\sin p_3 - \beta_6p_1,\\
 \tilde{\beta_6} &=& \beta_6, 		  
\end{array}
\end{equation} 
where $\alpha_i$ and $\beta_i$, $i=1, \ldots, 6$ are the arbitrary parameters that determine the factors in the product space $P$ (\ref{ps}) according to the general formula (\ref{gKt}). 
Recall that the group action $SE(2) \circlearrowright \cK^2(\Eset^2)$ yields  four types of non-trivial orbits (i.e., the orbits of dimesions 1, 2 and 3) corresponding to  the elements of $\cK^2(\Eset^2)$ with almost everywhere  pointwise distinct eigenvalues. As is well known, these four types of orbits correspond to four types of Killing two-tensors whose geometric and algebraic properties are equivalent up to the action of the isometry group $SE(2)$, namely the Killing tensors whose eigenvalues and eigenvectors generate elliptic-hyperbolic, polar, parabolic and cartesian coordinate systems respectively. Moreover, polar, parabolic and cartesian coordinate systems are all degeneracies of  elliptic-hyperbolic coordinate system, characterized by the existence of two foci, which are the singlular points where the eigenvalues of the corresponding Killing two-tensor coincide (see, for example,  \cite{MST02} and the relevant references therein for more details). In our setting, namely for the classification problem determined by the group action $SE(2)\circlearrowright P$, the analogue of the elliptic-hyperbolic coordinate system generated by a Killing two-tensor $K \in \cK^2(\Eset^2)$, is a pair  $(K_1, K_2) \in P$, where $K_1, K_2 \in  \cK^2(\Eset^2)$ are both of the elliptic-hyperbolic type. Moreover, the four foci of $K_1$ and $K_2$ are in the  general position forming an arbitrary quadrilateral (with no symmetry structure).  Let $S_1, S_2$ and $S_3, S_4$ be the foci of $K_1$ and $K_2$ respectively. Denote by $(x_i, y_i)_{S_i}$ their respective coordinates given in terms of the cooresponding arbirtrary vector space parameters $\alpha$'s and $\beta$'s (\ref{JA}). Then, 
\begin{equation}
\label{F1F2}
\begin{array}{lr}
(x_1,y_1)_{S_1}  = & \\
& \left(\frac{-\beta_5}{\beta_6} + \frac{1}{\beta_6}\left(\frac{\sqrt{\Delta'_6} - \sigma_1}{2}\right)^{1/2}, 
\frac{-\beta_4}{\beta_6} + \frac{1}{\beta_6}\left(\frac{\sqrt{\Delta'_6} + \sigma_1}{2}\right)^{1/2}\right), \\[1cm]
(x_2,y_2)_{S_2}  = & \\
& \left(\frac{-\beta_5}{\beta_6} - \frac{1}{\beta_6}\left(\frac{\sqrt{\Delta'_6} - \sigma_1}{2}\right)^{1/2}, 
\frac{-\beta_4}{\beta_6} - \frac{1}{\beta_6}\left(\frac{\sqrt{\Delta'_6} + \sigma_1}{2}\right)^{1/2}\right), 
\end{array}
\end{equation} 
\begin{equation}
\label{F3F4}
\begin{array}{lr}
(x_3,y_3)_{S_3}  = & \\
& \left(\frac{-\alpha_5}{\alpha_6} + \frac{1}{\alpha_6}\left(\frac{\sqrt{\Delta'_3} - \sigma_2}{2}\right)^{1/2}, 
\frac{-\alpha_4}{\alpha_6} + \frac{1}{\alpha_6}\left(\frac{\sqrt{\Delta'_3} + \sigma_2}{2}\right)^{1/2}\right), \\[1cm]
(x_4,y_4)_{S_4}  = & \\
& \left(\frac{-\alpha_5}{\alpha_6} - \frac{1}{\alpha_6}\left(\frac{\sqrt{\Delta'_3} - \sigma_2}{2}\right)^{1/2}, 
\frac{-\alpha_4}{\alpha_6} - \frac{1}{\alpha_6}\left(\frac{\sqrt{\Delta'_3} + \sigma_2}{2}\right)^{1/2}\right), 
\end{array}
\end{equation} 
where $\sigma_1 = \beta_4^2 -\beta_5^2 + \beta_6(\beta_2-\beta_1)$, $\sigma_2 = \alpha_4^2 -\alpha_5^2 + \alpha_6(\alpha_2-\alpha_1)$, while  $\Delta'_3$ and $\Delta'_6$ are given below by (\ref{I33}) \cite{MST02, AMS08}. 

According  to the Fundamental Theorem on invariants of regular Lie group action \cite{olver99} the action $SE(2)\circlearrowright P$ given by (\ref{JA}) yields $9 = \dim P - \dim SE(3)$ fundamental joint invariants. It was shown in \cite{AMS08} that the following 9 functions form a set of fundamental joint invariants of the group action (\ref{JA}): 
\begin{equation}
\label{I33}
\begin{array}{rcl}
{\Delta}'_1 & = & \alpha_6,\\[0.2cm]
{\Delta}'_2 & = & \alpha_6(\alpha_1 + \alpha_2) - \alpha_4^2 - \alpha_5^2,\\[0.2cm]
{\Delta}'_3 & = & (\alpha_6(\alpha_1 - \alpha_2) - \alpha_4^2 + \alpha_5^2)^2 + 4(\alpha_6 \alpha_3 + \alpha_4 \alpha_5)^2,\\[0.2cm]
{\Delta}'_4 & = & \beta_6,\\[0.2cm]
{\Delta}'_5 & = & \beta_6(\beta_1 + \beta_2) - \beta_4^2 - \beta_5^2,\\[0.2cm]
{\Delta}'_6 & = & (\beta_6(\beta_1 - \beta_2) - \beta_4^2 + \beta_5^2)^2 + 4(\beta_6 \beta_3 + \beta_4 \beta_5)^2, \\[0.2cm]
{\Delta}'_7 &= &d^2(S_2,S_3) = (x_2-x_3)^2 + (y_2-y_3)^2, \\[0.2cm]
{\Delta}'_8 &= &d^2(S_1,S_3) = (x_1-x_3)^2 + (y_1-y_3)^2, \\[0.2cm]
{\Delta}'_9 &=& d^2(S_2,S_4) = (x_2-x_4)^2 + (y_2-y_4)^2, 
\end{array}
\end{equation}
where $d^2(S_2, S_3)$, $d^2(S_1, S_3)$, $d^2(S_2, S_4)$ are three  joint ivariants representing the square distances between foci  of $K_1$ and $K_2$. Consider the resulting quadrilateral $S_1S_2S_3S_4$. Obviously, it completely determines the geometry of the corresponding pair $(K_1, K_2) \in P$. Moreover, one can fix it by fixing the interpoint distances $d(F_1, F_2)$, $d(S_3,S_4)$, $d(S_2, S_3)$, $d(S_1, S_3)$, $d(S_2, S_4)$, which are clearly invariants and joint invariants of the group action (\ref{JA}). Note that $d(S_1, S_2) = 2k_1$, $d(S_3, S_4) = 2k_2$, where $$k_1^2 = \frac{\sqrt{(\alpha_4^2-\alpha_5^2 + \alpha_6(\alpha_2-\alpha_1))^2 + 4(\alpha_6\alpha_3 + \alpha_4\alpha_5)^2 }}{\alpha_6},$$ $$k_2^2 = \frac{\sqrt{(\beta_4^2-\beta_5^2 + \beta_6(\beta_2-\beta_1))^2 + 4(\beta_6\beta_3 + \beta_4\beta_5)^2 }}{\beta_6}.$$ Recall also that
any joint invariant ${\cal J}(S_1,S_2,S_3, S_4)$ of the non-transitive action so defined can be written as a function of the  distances $d(S_1,S_2)$, $d(S_2,S_3)$, $d(S_3,S_4)$, $d(S_4,S_1)$ (this result is also known as ``the Weyl theorem on joint invariants'' \cite{olver99,weyl46}). 

Having constructed this groundwork, we can now proceed and use the language of the ITKT to invariantly characterize the Smorodinsky-Winternitz potential (\ref{sw}).

\section{Invariant characterization of the Smorodinsky-Winternitz potential}
\label{icsw}

As is well known the Smorodinsky-Winternitz potential of the Hamiltonian system defined by (\ref{sw}) is superintegrable. Furthermore, its superintegrability is afforded by first integrals which are quadratic in the momenta (\ref{fiq}), which follows immediately from the form of the SW potential.  Indeed, since the potential in question is in the form $V = f(x) + g(y)$ it follows that it is separable in (canonical) cartesian coordinates, which means that the Hamiltonian flow defined by (\ref{sw}) admits a first integral $F_1$ of the form: 

\begin{equation}
\label{f1}
F_1  = K^{ij}_1p_ip_j + U_1,\quad i,j = 1,2,
\end{equation} 
where $K_1^{11} = 1$, $K_1^{21} = K_1^{12} = K_1^{22} = 0$.

Next, since in the potential (\ref{sw}) is of the form $V(r,\theta) = \frac{f(\theta)}{r^2} + g(r)$ in polar coordinates, the Hamiltonian system defined by (\ref{sw}) is separable in (canonical) polar 
coordinates, which means that it admits the following quadratic first integral of motions: 
\begin{equation} 
\label{f2}
F_2  = K_2^{ij}p_ip_j + U_2,\quad i,j = 1,2,
\end{equation}
where the components of the corresponding Killing two-tensor are as follows:  $K_2^{11} = y^2$, $K_2^{21} = K_2^{12} = -xy$, $K_2^{22} = x^2$. It follows immediately that the Hamiltonian sysem defined by (\ref{sw}) is superintegrable (note $\{F_1, F_2\} \not=0$) and multi-separable. It also follows that the SW potential is separable in (canonical) elliptic-hyperbolic coordinates, which is a consequence of the fact that the function  $F_3 = F_1+F_2$ is a first integral of motion (see, for example, \cite{MST02, AMS08} for more details). Therefore, the Smorodinsky-Winternitz potential is separable in cartesian, polar and, - as a consequence, - in elliptic-hyperbolic coordinates. Separation in (canonical) polar and elliptic-hyperbolic coordinates implies that 
 in terms of the corresponding joint invariants (see Section \ref{itkt}) the quadrilateral $S_1S_2S_3S_4$ degenerates in this case. Let $S_1$ and $S_2$ be the  singular points (foci) of the Killing two-tensor $K_3$ of the first integral $F_3 = F_1 + F_2$, where $F_1$ and $F_2$ are given by (\ref{f1}) and (\ref{f2}) respectively.

 Next, let $S_3$ be the singular point of the Killing tensor defined by (\ref{f2}). Since, both $K_2$ and $K_3$ are in canonical form, we conclude that the quadrilateral $S_1S_2S_3S_4$ degenerates in this case, which is prompted by the invariant conditions $S_3=S_4$, $d(S_1,S_3) = d(S_2,S_3)$, or, in terms of the fundamental joint invariants (\ref{I33}),  we have  $\Delta'_3 = 0$, $\Delta'_8 = \Delta_9'$. 
 
Conversely, if we start with a general potential $V$ given by (\ref{h1}) and then impose that the corresponding Hamiltonian system admits two quadratic first integrals of the form (\ref{f1}) and (\ref{f2}) defined by Killing two-tensors $K_1$ and $K_2$ respectively, the corresponding compatibility conditions (\ref{cc}) given by $K_1$ and $K_2$  will yield the Smorodinsky-Winternitz potential (\ref{sw}) \cite{adlam05, FMSUW65}. 

\medskip 
{\em Conlcusion}: the superintegrable system of the Smorodinsky-Winternitz potential is  defined by a pair of Killing two-tensors $(K_2, K_3) \in P$, whose position in the orbit space $P/SE(2)$ is determined by the following invariant conditions: 
\medskip 

\begin{equation} \label{invsw}
\Delta'_1 \not=0, \, \Delta'_3 = 0,\, \Delta'_4 \not= 0, \, \Delta'_6 \not=0, \, \Delta'_7 = \Delta'_8 = \Delta'_9.
\end{equation}
Hence, we have 
 \begin{theorem}
\label{t52}
Let the potential $V$ of the general Hamiltonian (\ref{h1}) be compatible via (\ref{cc}) with a pair of  
Killing two-tensors  $(K_1, K_2) \in P$. Then the following statements are equivalent:  
\begin{itemize}
\item[(1)] The pair $(K_1, K_2) \in P$ is invariantly characterized by the conditions (\ref{invsw}). 

\item[(2)] The potential $V$ given by (\ref{h1}) is the Smorodinsky-Winternitz potential (\ref{sw}).  
\end{itemize}
\end{theorem}

That is, by invariantly characterizing the pair $(K_2, K_3) \in P$, we have thus characterized the Smorodinsky-Winternitz potential that they define. 

Now let us investigate what happens to the SW potential (\ref{sw})  if we weaken the invariant conditions (\ref{invsw}). More precisely, let us first suppose that the Hamiltonian system defined by the
natural Hamiltonian (\ref{sw}) admits  two quadratic first integrals of motion

\begin{equation}
\label{f3}
F_1  = K^{ij}_1p_ip_j + U_1,\quad i,j = 1,2,
\end{equation} 
and
\begin{equation} 
\label{f4}
F_2  = K_2^{ij}p_ip_j + U_2,\quad i,j = 1,2,
\end{equation}
Furthermore, we will assume that the Killing tensor $K_1$ of (\ref{f3}) generates polar coordinates, while $K_2$ of (\ref{f4}) generates elliptic-hyperbolic and they are are in general position, which means their respective singular points $S_3$ (of $K_1$) and $S_1$, $S_2$  (of $K_2$) form a general triangle $\triangle S_1S_2S_3$. This arrangement entails the following invariant conditions in terms of the fundamental joint invariants (\ref{I33}) for the pair $(K_1, K_2) \in P$: 
\begin{equation}
\Delta'_1 \not=0, \, \Delta'_3 = 0,\, \Delta'_4 \not= 0, \, \Delta'_6 \not=0, \, \Delta'_7 = \Delta'_9 \not= \Delta'_8.
\end{equation}
Without loss of generality we can assume that $K_2$ is in canonical form, while $K_1$ is not, which means that the respective components of the two Killing tensors are as follows: 
\begin{equation}\label{gp}
K_1^{11} = (y-b)^2, K_1^{21} = K_1^{12} = -(x-a)(y-b), K_2^{22} = (x-a)^2, \, a, b \in \mathbb{R},
\end{equation}
and 
\begin{equation}\label{ceh}
K_2^{11} = y^2 + \ell, K_2^{21} = K_2^{12} = -xy, K_2^{22} = x^2, \, k \in \mathbb{R}, 
\end{equation}
where the parameter $\ell = 2k^2$ is an invariant determined by the condition $$k^2 = \frac{\sqrt{(\beta_4^2-\beta_5^2 + \beta_6(\beta_2-\beta_1))^2 + 4(\beta_6\beta_3 + \beta_4\beta_5)^2 }}{\beta_6} = \frac{\sqrt{(\Delta'_6)^2}}{\Delta'_4}.$$ 

\medskip
{\em Question}: How does this setting affect the form of the Smorodinsky-Winternitz potential (\ref{sw})? 
\medskip

Our first observation is that since $F_1$ and $F_2$ are the first integrals of the Hamiltonian system defined by (\ref{sw}), the Killing two-tensor given by the linear combination
\begin{equation}
\label{Kg}
K_g = c_1K_1 + c_2K_2 + c_3g, \, c_1, c_2, c_3 \in \mathbb{R}
\end{equation}
must also be compatible with the SW potential $V$ given by (\ref{sw}), i.e., $\mbox{d}(\hat{K}_gdV) = 0$. However, $K_g$ depends on six  arbitrary parameters, namely $\ell$, $a$, $b$, $c_1$, $c_2$, $c_3$, but recall that 6 is exactly the dimension of  $\dim \cK^2(\Eset^2)$. This is only possible if $K_g$ (\ref{Kg}) is given by the general formula (\ref{gKt}), which immediately implies that the potential (\ref{sw}) must be  trivial, i.e., $\omega = \alpha  = \beta$. 
Therefore the  only parameters that can be set to  zero are $a$ and $b$. Having discussed the case 
\begin{itemize}
\item[1)] $a \not= 0, b \not=0$,
\end{itemize}
which reduced the SW potential (\ref{sw}) to a trivial one, now let us consider the following remaining three cases:

\begin{itemize}
\item[2)] $a = 0, b \not=0$,

\item[3)] $a\not = 0, b = 0$, 

\item[4)] $a = b = 0.$
\end{itemize}

First,  we note that in terms of the fundamental joint invariants (\ref{I33}) these cases can be invariantly characterized as follows: 

\begin{itemize}
\item[1)] $\Delta'_1 \not=0, \, \Delta'_3 = 0,\, \Delta'_4 \not= 0, \, \Delta'_6\not=0, \,  \Delta'_7\Delta'_8\Delta'_9\not=0, \, \Delta'_7 \not= \Delta'_8\not=  \Delta'_9$, 

\item[2)] $\Delta'_1 \not=0, \, \Delta'_3 = 0,\, \Delta'_4 \not= 0, \, \Delta'_6\not=0, \,  \mbox{the area of} \, \triangle S_1S_2S_3 = 0.$
Note that in this case the singular points $S_1, S_2, S_3$ are on the $x$-axis. The area of $\triangle S_1S_2S_3$ is obviously also a joint invariant. 

\item[3)] $\Delta'_1 \not=0, \, \Delta'_3 = 0,\, \Delta'_4 \not= 0, \, \Delta'_6\not=0, \, \Delta'_7 - \Delta_8' =0.$ Note in this case $\triangle S_1S_2S_3$ is an isosceles triangle. 

\item[4)] $\Delta'_1 \not=0, \, \Delta'_3 = 0,\, \Delta'_4 \not= 0, \, \Delta'_6 \not=0, \, \Delta'_7 = \Delta'_9 \not= \Delta'_8.$

\end{itemize}

Next, we note that the arbitrary parmeters $a$ and $b$ are obviously also joint invariants given in terms of the fundamental joint invariants (\ref{I33}) by the following formulas: 

\begin{equation}
\label{a}
a = \frac{(\Delta'_4)^2(\Delta'_8 - \Delta'_7) - \frac{1}{2}\Delta'_6}{2\Delta'_4\sqrt{\Delta'_6}}
\end{equation}
and
\begin{equation}
b = \sqrt{\Delta'_7 - \left(a - \frac{\sqrt{\Delta'_6}}{2\Delta'_4}\right)^2}.
\end{equation}
Solving the PDEs defined by the compatibility condition (\ref{cc}) for the Killing two-tensor (\ref{gp}) and the SW potential (\ref{sw}), we arrive at the following equation
\begin{equation}
\omega^2(bx - ay) + \frac{a(b-y)\alpha}{x^4} + \frac{b(- a + x)\beta}{y^4} = 0,
\end{equation}
which puts in evidence that 
\begin{itemize}
\item[2)] $a = 0, b \not=0$ yields the  potential $V = \frac{\beta}{y^2}$,

\item[3)] $a \not= 0, b = 0$ yields the  potential $V = \frac{\alpha}{x^2}$, 

\item[4)] $a =0$, $b =0$ yields the general Smorodinsky-Winternitz potential (\ref{sw}). 
\end{itemize} 

We conclude therefore that the more geometric structure the pair $(K_1, K_2) \in P$ has, the less general the Smorodinsky-Winterniz potential (\ref{sw}) becomes. We also note that in each case described above  the corresponding pair $(K_1, K_2) \in P$ belongs to a different orbit of the group action $SE(2) \circlearrowright P$ (\ref{JA}). Indeed, in the above the corresponding orbits have been distinquished by means of joint invariants (\ref{I33}) of the action (\ref{JA}).  Extending in this paper the results obtained in \cite{AMS08}, we have demonstrated that the invariant classification of the orbits $P/SE(2)$  leads to a classification of the corresponding superintegrable  potentials defined by quadratic first integrals in $\Eset^2$. 

\section{A remark on the Trembley-Turbiner-Winternitz potential}

Consider now the Trembley-Turbiner-Winternitz (TTW)  potential defined by the natural Hamiltonian (\ref{ttw}). Clearly, it is separable in (canonical) polar coordinates, admitting a first integral of motion defined by the Killing two-tensor  (\ref{gp}) for $a= b = 0$. 

\medskip 
{\em Question}: For what value(s) of the parameter $k$  does the TTW potential (\ref{ttw}) admit an additional  functionally independent quadratic first integral of motion? 
\medskip 

Assuming $k$ is real, we cannot transform the TTW system back into cartesian coordinates and so we must carry out all our calculations in terms of polar coordinates.   Hence,  we first rewrite the general formula the vector space 
(\ref{gKt}) in terms of polar coordinates  as follows: 
\begin{equation} \label{gKtpolar}
K_g =   A^{i j }X_i \odot X_j + B^i X_i \odot R + \beta_6 R, \, i,j = 1,2,
\end{equation} 
where $\odot$ is the symmetric tensor product and the generators $X_1, X_2, R$  of the Lie algebra {\gothic se}(2)   are given by 

\begin{eqnarray}
X_1 & = &  \cos\theta \frac{\partial}{\partial r} - \frac{\sin\theta}{r}\frac{\partial}{\partial \theta},\\
X_2 & = & \sin\theta\frac{\partial}{\partial r} + \frac{\cos\theta}{r}\frac{\partial}{\partial \theta},\\
R  &  = & \frac{\partial}{\partial \theta}. 
\end{eqnarray}
Here \begin{equation}
A^{i j} = \left( \begin{array}{cc}
\beta_1 & \beta_3\\
\beta_3 & \beta_2
\end{array} \right),\quad
B^i = \left(\begin{array}{c}\beta_4\\ \beta_5 \end{array} \right). 
\label{eq:parameters}
\end{equation}
In view of the above  the components of the general Killing tensors defining $\cK^2(\Eset^2)$ in polar coordinates are given by 
\begin{eqnarray}
K^{11} &= &
 2 \beta_3 \cos{\theta } \sin{\theta}+ \beta_1 \cos{\theta}^2 + \beta_2\sin{\theta}^2,    \\
K^{1 2}  &=&  \frac{2 \beta_3  \cos{2 \theta}}{r}-\frac{2\beta_1 \cos{\theta } \sin{\theta } }{r}+\frac{2\beta_2 \cos{\theta } \sin{\theta } }{r}
 +2\beta_4 \cos{\theta } +2 \beta_5 \sin{\theta },   \label{eq:polarKT} \\
K^{2 2} &=&  \beta_6 -\frac{2 \beta_3  \cos{\theta } \sin{\theta }}{r^2}+\frac{\beta_1\sin{\theta }^2 }{r^2}+\frac{\beta_2\cos{\theta }^2}{r^2}-\frac{2\beta_4 \sin{\theta } }{r}+\frac{2 \beta_5 \cos{\theta } }{r}.  
\end{eqnarray} 

Substituting (\ref{gKtpolar}) and the potential of the natural Hamiltonian (\ref{ttw}) into the compatibility condition (\ref{cc}), we arrive at a complicated trigonometric equation in terms 
of the parameters $k$, $\alpha$, $\beta$, $\omega$, $\beta_i$, $i= 1, \ldots, 6$. To solve it, we expand the equation in terms of the following set $M$ of trigonometric functions
 \begin{equation}
\begin{split}
M = \{1, &\cos(\theta - 2mk), \sin(\theta-2mk),\cos(\theta+2mk),\sin(\theta+2mk),\cos(2\theta - 2mk), \\
&\sin(2\theta-2mk),\cos(2\theta+2mk),\sin(2\theta+2mk)\}, \quad m=0,\dots, 4.
\end{split}
\label{trigbasis}
\end{equation}
The set $M$ is linearly dependent when the arguments of the trigonometric  functions coincide, which depends on the values assumed by the parameter $k$.  Indeed, we have that the set (\ref{trigbasis})  is linearly \emph{dependent} for the following values of $k$: 
\begin{equation}
\begin{split}
k = &\pm 2,\pm 1\frac{1}{2},\pm 1,\pm \frac{1}{2},\pm \frac{1}{4},\pm \frac{1}{6},\pm \frac{1}{8},\pm \frac{1}{10},\pm \frac{1}{12},\pm \frac{1}{14},\pm \frac{1}{16},\pm \frac{3}{4},\pm \frac{2}{3},\\\
&\pm\frac{3}{8},\pm \frac{1}{3},\pm \frac{3}{10},\pm \frac{2}{7},\pm \frac{3}{14},\pm \frac{1}{5},\pm \frac{3}{16},\pm \frac{2}{5},\pm \frac{1}{7}.
\end{split}
\label{special}
\end{equation}
In the case when $k$ does not equal one of the above values, the set $M$ is  linearly independent and so we can set each of the coefficients to zero in the trigonometric equations that we obtained upon the substitution of (\ref{gKtpolar}) and (\ref{ttw}) into (\ref{cc}).   This argument leads to an overdetermined system of polynomial equations which we can solve for the variables 
$k$, $\alpha$, $\beta$, $\omega$, $\beta_i$, $i= 1, \ldots, 6$. Though, the problem at this point can be simplified according to the following argument. Since the TTW potential (\ref{ttw}) is separable in polar coordinates, if it were to admit another quadratic first integral of motion, it would have been defined by a Killing two-tensor which is an element of the 5-dimensional vector subspace 
\begin{equation}
\label{reduced}
K_s =   A^{i j }X_i \odot X_j + B^i X_i \odot R, \, i,j = 1,2,
\end{equation}
i.e., the vector space (\ref{gKtpolar}) with the ``polar" term factored out. Substituting this Killing two-tensor $K_s$ into the compatibility condition (\ref{cc}) for the SW potential (\ref{ttw}), we arrive at a system of  equations, which is possible to expand over the basis  $L = \{1,\cos\theta,\sin\theta,\cos2\theta,\sin2\theta\},$ which results in the following restriction on the parameters of the Killing two-tensor (\ref{reduced}): 
$$\beta^2_4 + \beta^2_5 = 0.$$
Hence, we conclude that if the potential (\ref{ttw}) is separable in a coordinate system, other than polar, it can only be the cartesian system of coordinates. Now we can repeat the same argument, but for a much simpler Killing two-tensor, namely the Killing two-tensor that defines  cartesian coordinates in general position (not canonical) \cite{MST02}, namely: 
\begin{equation}
K_c= 
\left(
\begin{array}{cc}
 \cos^2(\theta -\phi ) & -\frac{1}{2} r \sin(2 (\theta -\phi )) \\
 -\frac{1}{2} r \sin(2 (\theta -\phi )) & r^2 \sin^2(\theta -\phi)
\end{array}
\right),
\label{cartesian}
\end{equation}
where $\phi$ denotes the angle by which the cartesian coordinate system is rotated.  Substituting (\ref{cartesian}) and (\ref{ttw}) into the compatibility condition (\ref{cc}), we get the following equation: 
\begin{equation}
\begin{split}
&-4 k \beta \cos^2(\theta -\phi ) \cot(k \theta) \csc^2(k \theta)+4 k \beta \cot(k \theta) \csc^2(k \theta) \sin^2(\theta -\phi )\\
&-4 \beta \csc^2(k \theta) \sin(2 (\theta -\phi ))+2 k^2 \beta \cot^2(k \theta) \csc^2(k \theta) \sin(2 (\theta -\phi ))\\
&+k^2 \beta \csc^4(k \theta) \sin(2 (\theta -\phi ))-4 \beta \sec^2(k \theta) \sin(2 (\theta -\phi ))\\
&+k^2 \alpha \sec^4(k \theta) \sin(2 (\theta -\phi ))+2 k^2 \alpha \sec^2(k \theta) \sin(2 (\theta-\phi)) \tan^2(k \theta)\\
&+\cos(2 (\theta -\phi )) \left(-2 k \beta \cot(k \theta) \csc^2(k \theta)+6 k \alpha \sec^2(k \theta) \tan(k \theta )\right) = 0.
\end{split}
\label{cartKT}
\end{equation}
We can then expand this over a subset $N \subset M$ given by 
\begin{equation}
N = \{1, \cos(2(1 \pm k)\theta), \sin(2(1 \pm k)\theta)\}, \quad m=1,\dots,3,
\label{eq:trig-basis-subset}
\end{equation}  
which again is linearly dependent for the following values of $k$
\begin{equation}
k = \pm 1,\pm 2,\pm \frac{2}{3},\pm \frac{1}{2},\pm \frac{2}{5},
\label{eq:reduced-k}
\end{equation}
which are indeed a subset of the values obtained in (\ref{special}).  Expanding the equation (\ref{cartKT}) over the set $N$, and assuming it to be linearly independent (i.e. excluding the values of $k$ given by (\ref{eq:reduced-k})), we arrive at the follwing system of polynomial equations in the variables $\phi, k, \alpha,\beta$, namely
\begin{align*}
&\left(-1+4 k^2\right) (\beta+\alpha )  \cos\phi \sin\phi =0,\\
&(-2-k) (1+k) (\beta-\alpha ) \cos\phi \sin\phi=0,\\
&(1+k) (1+2 k) (\beta +\alpha ) \cos\phi \sin\phi=0,\\
&(-2+k (15+23 k)) (\beta -\alpha ) \cos\phi \sin\phi=0,\\
&(-2+k (-15+23 k)) (\beta -\alpha ) \cos\phi \sin\phi=0,\\
&(-1+k) (-1+2 k) (\beta +\alpha ) \cos\phi \sin\phi=0,\\
&(-2+k) (-1+k) (\beta-\alpha ) \cos\phi \sin\phi=0,\\
&8 \left(-1+4 k^2\right) (\beta +\alpha )\cos^2\phi-8 \left(-1+4 k^2\right) (\beta +\alpha ) \sin^2\phi =0,\\
&(1+k) (2+k) (\beta-\alpha) \cos^2\phi+(-1-k) (2+k) (\beta -\alpha ) \sin^2\phi=0,\\
\end{align*}
\begin{equation*}
\begin{split}
&4 \beta  \cos^2\phi+12 k \beta \cos^2\phi+8 k^2 \beta \cos^2\phi+4 \alpha  \cos^2\phi+12 k \alpha  \cos^2\phi\\
&+8 k^2 \alpha  \cos^2\phi-4 \beta  \sin^2\phi-12 k \beta \sin^2\phi-8 k^2 \beta  \sin^2\phi-4 \alpha \sin^2\phi\\
&-12 k \alpha  \sin^2\phi-8 k^2 \alpha  \sin^2\phi=0,\\
&-2 \beta  \cos^2\phi-15 k \beta  \cos^2\phi+23 k^2 \beta \cos^2\phi+2 \alpha  \cos^2\phi+15 k \alpha  \cos^2\phi\\
&-23 k^2 \alpha \cos^2\phi+2 \beta\sin^2\phi+15 k \beta \sin^2\phi-23 k^2 \beta \sin^2\phi-2 \alpha \sin^2\phi\\
&-15 k \alpha  \sin^2\phi+23 k^2 \alpha  \sin^2\phi=0,\\
&4 \beta  \cos^2\phi-12 k \beta  \cos^2\phi+8 k^2 \beta \cos^2\phi+4 \alpha  \cos^2\phi-12 k \alpha \cos^2\phi\\
&+8 k^2 \alpha \cos^2\phi-4 \beta  \sin^2\phi+12 k \beta \sin^2\phi-8 k^2\beta \sin^2\phi-4 \alpha \sin^2\phi\\
&+12 k \alpha  \sin^2\phi-8 k^2 \alpha \sin^2\phi=0,\\
&2 \beta  \cos^2\phi-3 k \beta \cos^2\phi+k^2 \beta \cos^2\phi-2 \alpha \cos^2\phi+3 k \alpha\cos^2\phi\\
&-k^2 \alpha  \cos^2\phi-2\beta \sin^2\phi+3 k \beta \sin^2\phi-k^2 \beta \sin^2\phi+2 \alpha  \sin^2\phi\\
&-3 k \alpha  \sin^2\phi+k^2 \alpha  \sin^2\phi=0,\\
&(-2+k (15+23 k)) (\beta -\alpha ) \cos^2\phi+(2-k (15+23 k)) (\beta-\alpha ) \sin^2\phi =0.
\end{split}
\end{equation*}
There is no non-trivial solution for which the above system is satisfied.  Considering the set of linearly dependent functions, we substitute into equation (\ref{cartKT}) with the special values of $k$ given in (\ref{eq:reduced-k}) to yield the following possible cases

\begin{itemize}
	\item[1)] $k = \pm 1$, \\
  $\phi = 0$, which implies the polar coordinate system is in canonical form; 
	\item[2)] $k =\pm 2,\pm 2/3,\pm 1/2,\pm 2/5 $,\\
	$\omega=\alpha=\beta = 0$, i.e., $V$ is a trivial potential. 
\end{itemize}
Thus, we conclude that $k = \pm 1$ are the only values of $k$ for which the TTW potential  (\ref{ttw}) admits two quadratic first integrals of motion. For these values of $k$ it reduces to the SW potential (\ref{sw}). Hence, we have proven the following
\begin{proposition}
The TTW potential given by (\ref{ttw}) is a multi-separable superintegrale potential iff $k=\pm 1$. 
\end{proposition}

\section{Conclusions} 
We have invariantly characterized the Smorodinsky-Winternitz superintegrable potential modulo the action of the isometry group $SE(2)$  in terms of the joint invariants of the associated Killing two-tensors by treating them as points in the product space (\ref{ps}). This result demonstrates that it can be extended to the problem of classification of arbitrary superintegrable potentials defined in spaces of constant curvature. More specifically, such a classification can be carried out by classifying first any associated Killing tensors that define first integrals of motion of the class of superintegrable systems in question (e.g., the superintegrable systems that admit only quadratic in the momenta first integrals of motion). Such Killing tensors ought to be considered as elements of the corresponding product spaces, whose factors are the corresponding vector spaces of Killing tensors representing the corresponding orbits in the orbit space induced by the action of the isometry group in such product spaces. 

In addition, we have proven that the Tremblay-Turbiner-Winternitz potential is a superintegrable system with two functionally independent quadratic first integrals of motion only  for $k=\pm 1$, that is when it reduces to the Smorodinsky-Winternitz potential. 

\bigskip 

\textbf{Acknowledgements}. The first author wishes to thank the organizers of the symposium on ``Superintegrability, Exact Solvability, and Special Functions'' (Cuernavaca, Mexico, February 20-24, 2012) for the invitation and also acknowledges useful discussions with Willard Miller, Jr. and Sara Post on the subject of the Tremblay-Turbiner-Winternitz potential. The research was supported in part by  NSERC in the form of a Discovery Grant (RGS). This material is based upon work supported by the National Science Foundation Graduate 
Research Fellowship under Grant No. 1048093 (AY). Any opinion, findings, and 
conclusions or recommendations expressed in this material are those of the authors(s) and do not 
necessarily reflect the views of the National Science Foundation.  All of the necessary computations have been performed with the aid of the {\em Mathematica} computer algebra package.

\LastPageEnding

\end{document}